\documentclass[aps,prl,twocolumn,showpacs, superscriptaddress, groupedaddress]{revtex4-1}

\usepackage{graphicx}
\usepackage{amsmath}
\usepackage{amssymb}
\usepackage[colorlinks,citecolor=blue,linkcolor=blue]{hyperref}
\usepackage[normalem]{ulem}
\usepackage{soul}

\newcommand{\angstrom}{\mbox{\normalfont\AA}}
\usepackage{array}
\newcolumntype{C}[1]{>{\centering\let\newline\\\arraybackslash\hspace{0pt}}m{#1}}

\newcommand{\change}[1]{#1}

\newcommand{\latestChange}[1]{#1}
\newcommand{\atLast}[1]{#1} 
\newcommand{\again}[1]{#1} 
\newcommand{\modify}[1]{{#1}}
\newcommand{\finalRevision}[1]{{#1}}

\begin{document}

\title{Floating potential of emitting surfaces in plasmas with respect to the space potential}
\author{B. F. Kraus}
\affiliation{Department of Astrophysical Sciences, Princeton University, Princeton, New Jersey, 08544 USA}
\author{Y. Raitses}
\affiliation{Princeton Plasma Physics Laboratory, Princeton University, Princeton, New Jersey, 08543 USA}
\date{\today}

\begin{abstract}
The potential difference between a floating emitting surface and the plasma surrounding it has been described by several sheath models, including the space-charge-limited sheath, the electron sheath with high emission current, and the inverse sheath \again{produced by} charge-exchange ion trapping. Our measurements reveal that each of these models has its own regime of validity. We determine the \again{potential} of an emissive filament \again{relative to the plasma potential}, emphasizing variations in emitted current density and neutral particle density. The \again{potential} of a filament in a diffuse plasma is first shown to vanish, consistent with the electron sheath model and increasing electron emission. In a denser plasma with ample neutral pressure, the \again{floating filament potential is positive}, as predicted by a derived ion trapping condition. Lastly, the filament floated negatively in a third plasma, where flowing ions and electrons and nonnegligible electric fields may have disrupted ion trapping. Depending on the regime chosen, emitting surfaces can float positively or negatively with respect to the plasma potential.
\end{abstract}

\maketitle

Any solid surface in contact with a plasma is surrounded by the sheath, a potential structure that controls particle and energy transport between the plasma and the surface \cite{Hershkowitz2005}. Sheath structure is complicated when the surface emits an electron current, which can be caused by impinging radiation or plasma particles. Emissive sheaths are present in divertors \cite{Harbour} and scrape-off layers \cite{Stangeby} in magnetic fusion devices, around dust grains in laboratory \cite{Samarian} and astrophysical \cite{Horanyi} plasmas, \change{around satellites \cite{Whipple}}, in RF plasma processing devices \cite{Anders} and around plasma probes \cite{SheehanPRL}. In all of these cases, the interplay between emitted and background plasmas determines the structure of sheath that forms. Predicting which structure exists is essential for understanding the heat and charge flux to the surface.

\begin{figure}[b]
\centering
\includegraphics[width=0.5\textwidth]{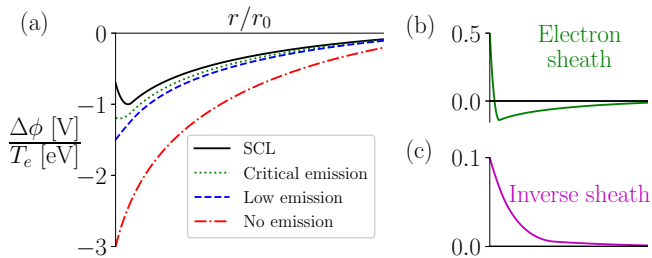}
\caption{\label{fig:potentialProfileCartoon}A series of potential profile cartoons, representative of the emissive sheath models from the text. The parameter $\Delta \phi$ is here defined as $\phi(r) - \phi_P$, where $\phi_P$ is the plasma potential far from the surface. Potential magnitudes are approximate but have realistic signs and relative sizes.}
\end{figure}

A surface emits a normalized current $\hat{J} = J_\text{emit}/J_e$ in a background electron current $J_e$. When $\hat{J}=0$, \again{mobile plasma electrons charge the surface negatively so that its potential is negative with respect to the plasma potential $\phi_P$ (in this work, all surface potentials called positive or negative are referenced to $\phi_P$).} The sheath is monotonic (see Fig.~\ref{fig:potentialProfileCartoon}a), electrically and thermally insulating the surface from plasma electrons. The sheath potential drop is weakened but still monotonic if $\hat{J} \ll 1$. At a critical emission near $\hat{J} \approx 1$, the electric field at the surface passes through zero; beyond this degree of emission, a potential well forms that restricts emitted electrons from reaching the bulk plasma. This state of sheath is termed space-charge-limited (SCL) \cite{HobbsWesson}. The strong emission greatly reduces the SCL sheath potential drop, but the surface is still negative and thus weakly insulated.

The above theory for \modify{$\hat{J} > 1$} has recently been modified by additional physical mechanisms, each of which reduces or erases the insulating negative potential of the SCL sheath. When object scale lengths are small compared to the plasma Debye length $\lambda_D$ or (for magnetized plasmas) the Larmor radius $\rho_L$, emitted electrons experience orbital motion effects \cite{Laframboise}, which lengthen the trajectories of trapped emitted electrons between emission and surface reabsorption. These two-dimensional effects have been shown to not only reduce the magnitude of the potential dip \cite{Robertson}, but also to build up the potential of floating emitting surfaces \cite{ChenX2017, Fruchtman, Cavalier, Taccogna}. When electron emission from a small object sufficiently overwhelms the incoming plasma current, much of the sheath remains negative, but the surface floats above $\phi_P$ \cite{Delzanno} as in Fig.~\ref{fig:potentialProfileCartoon}b. This is the nonmonotonic electron sheath. Another mechanism---ion trapping inside the SCL potential dip---leads to the same consequence. In this case, ambient neutrals charge-exchange near the object and are often slow enough to become trapped in the potential well \cite{Goree}. This new population gradually increases the surface potential up to or above $\phi_P$ \cite{Campanell2016}. As shown in Fig.~\ref{fig:potentialProfileCartoon}c, ion trapping can lead to a \again{monotonic and} positive potential profile, the so-called inverse sheath. Some similar mechanisms even affect the sheath potential for non-emitting surfaces, such as magnetic field geometry \cite{Beilis1997, Beilis1998} or charge-exchange ions trapped in effective potential wells due to geometry, not emitted space charge \cite{Lampe2001}. Each of these effects modifies the insulating potential sheath, influencing the electron flux to the surface.

The mechanisms that alter SCL sheaths coexist in many plasma environments, making their consequences difficult to isolate and experimentally validate. In this work, we present evidence from three plasma scenarios without magnetic fields where different sheath structures are accurately described by either extreme emission electron sheaths, ion trapping in the SCL dip, or standard SCL theory. \atLast{Measurements were made as follows: (1) In a low pressure discharge (discharge voltage \mbox{$V_D =$ 45 V}, current \mbox{$I_D =$ 1.37 A}) with a thermionic hollow cathode \mbox{\cite{Vekselman}}, the filament \modify{floated} negatively or at $\phi_P$; (2) Inside a 10-cm-diameter, 50--60 W ferromagnetic inductively coupled plasma source \mbox{\cite{GodyakSources, GodyakIEPC}}, the filament \again{floated} positively; (3) Placed several cm from the anode in the plume of a 2.6-cm-diameter, unmagnetized Hall thruster (\mbox{$V_D =$ 50 V}, \mbox{$I_D =$ 1.37 A}) with flowing electrons, ions, and neutrals \mbox{\cite{Fedotov, Smirnov2007}}, filaments remained negative.} The studied plasmas are stable, homogeneous across measurement scale lengths, and well-characterized by Langmuir probe measurements. In each scenario, we compare the emissive floating potential of a plasma-immersed surface to the plasma potential measured by sweep-biasing the same surface, thus inferring the magnitude of the emissive sheath potential drop at a single location. The sheath potential drop is not as descriptive as the full spatial profile of the sheath potential (a profile notoriously difficult to measure noninvasively), but it does provide strong evidence for transitions between sheath types. These results show that the sheath model can change based on plasma parameters, and that several physical models predict sheath transitions consistently with experiment.

\begin{figure}
\centering
\includegraphics[width=0.5\textwidth]{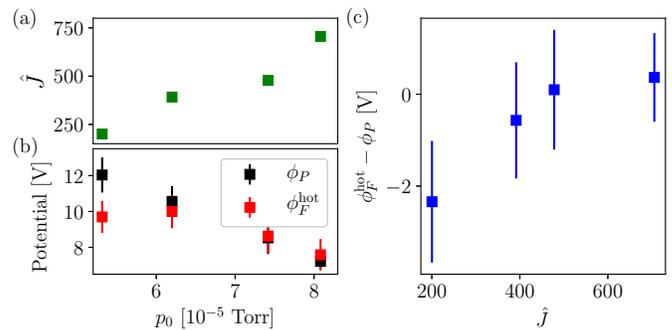}
\caption{\label{fig:noFlow} Four plasma conditions with varied neutral pressure $p_0$ are shown, each with monotonic electron energy distributions and low plasma density of $n_e \approx$ 3--5 $\times 10^{15}$ m$^{-3}$. Plot (a) shows \finalRevision{approximate} $\hat{J}$ from a filament immersed in these plasmas, while (b) compares the electric potential of the plasma, $\phi_P$, with the floating potential of a strongly emitting filament, $\phi_F^\text{hot}$. The potential difference $\phi_F^\text{hot} - \phi_P$ is shown versus $\hat{J}$ in (c).}
\end{figure}

A small object with radius $r_0 \lesssim \lambda_D$ submerged in a low pressure plasma can easily be heated so that $\hat{J} \gg 1$. For instance, a thoriated tungsten filament at \finalRevision{1900 K (2000 K) would emit a modest thermionic emission of $J_\text{emit} =$ 1.4 (3.4) A/cm$^2$,} as described by the Richardson-Dushman equation \cite{Dushman}; here \finalRevision{the surface temperature is approximated by the measured filament resistance and corroborated} by measuring the current to a negatively biased emitting surface. In contrast, a diode glow discharge in the plume of a hollow cathode produces a flux of $J_e = 0.02$ A/cm$^2$ (measured by a probe \modify{biased to $\phi_P$}) when plasma electron density and temperature are  $n_e = 3.3 \times 10^{15}$ m$^{-3}$ and $T_e^\text{eff} = 5.1$ eV, respectively (parameters were derived from integrals of the electron energy distribution function \cite{Godyak2011}). Such plasma parameters are typical for diffuse gas discharges. The current ratio $\hat{J}$ for this filament immersed in the described plasma are shown in Fig.~\ref{fig:noFlow}a, with currents \finalRevision{\mbox{$\hat{J} \approx$ 200--800} measured in a discharge where neutral pressure $p_0$ was varied. Note that the exponential dependence on wire temperature of $\hat{J}$ leads to large absolute uncertainties; however, relative emission current values compared between scenarios are discernible due to reproducible changes in filament resistance}. Increasing pressure causes the plasma electron temperature to decrease, so that plasma current to the surface falls and $\hat{J}$ increases.

The emissive floating potentials $\phi_F^\text{hot}$ of a filament and the colocated plasma potentials $\phi_P$ are shown in Fig.~\ref{fig:noFlow}b. These potentials and all others in this work were measured with respect to the grounded chamber walls, and $\phi_P$ was derived from Langmuir probe measurements as described in \cite{Godyak2011}. \change{Plasma conditions were free of significant voltage oscillations, and probe characteristics indicate ample resolution for potential differences to be detectable \cite{GodyakAlexandrovich}; the emissive filament was Ohmically heated on a 50\% duty cycle, and data was recorded only when no external voltage was applied \cite{Iizuka}.} Initially at \finalRevision{$\hat{J} \approx 250$}, the emissive filament \again{reached a negative potential of} \change{about \mbox{2 V $\approx T_e/$4,}} in agreement with the SCL model: $\phi_F^\text{hot} < \phi_P$. As the emitted current increased, however, the emitted flux of cold electrons near the object overwhelmed the incoming plasma flux, so that $\phi_F^\text{hot} \approx \phi_P$. This trend agrees quantitatively with OML$^+$ theory \cite{Delzanno}, which predicts the disappearance of the potential drop between plasma and surface at \finalRevision{$\hat{J} \approx$ 100--200}. This rough agreement is expected, since the OML$^+$ theory is general and the relative object sizes are comparable, with \mbox{$r_0/\lambda_D = 1$} in in the particle-in-cell (PIC) calculations performed in \cite{Delzanno} and \mbox{$r_0/\lambda_D$ = 0.3} in the present experiment. Although we cannot resolve the nonmonotonic electron sheath in space, the measured reduction in sheath potential is consistent with such a model, with surface emission overwhelming the incoming plasma current.

The sheath potential profile is not known; it is only constrained to be loosely positive in sign, a quality shared by the inverse sheath. As the above plasma conditions differ in ambient neutral pressure, we must check whether the reduction of the emitting sheath potential drop could be a neutral pressure effect. Indeed, the higher the neutral density in the potential dip of the SCL sheath, the more charge exchange events produce cold ions there. If these new ions are trapped in the SCL potential well, and if they diffuse out of the trap slowly enough to accumulate there, then their positive contribution to the space charge will gradually increase \cite{Campanell2016}. Because the trapped ions neutralize the negative space charge near the surface, the sheath potential drop is reduced. The importance of this effect is governed by how large the trapped ion density eventually grows.

\begin{figure}[t]
\centering
\includegraphics[width=0.4\textwidth]{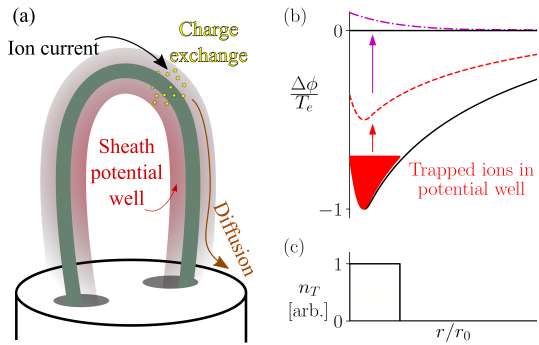}
\caption{\label{fig:cartoon} A cartoon (a) shows the accumulation and loss mechanisms of trapped ions inside an SCL potential well. The potential profiles (b) illustrate the gradual reduction of the sheath potential drop due to trapped ions, which are modeled with a toy density profile $n_T(r) = n_T$ for \mbox{$r_0 < r < r_0+W$} (c).}
\end{figure}

Measuring the density and motion of trapped ions is beyond \latestChange{the scope of this work}, but a heuristic model can predict their influence on the sheath. Informed by studies of ion trapping near biased plates \cite{ForestHershkowitz}, we model accumulation of charge-exchange ions in the SCL potential dip and their subsequent diffusion out of the trap. The geometry is cylindrical as shown in Fig.~\ref{fig:cartoon}a, with a filament of length $L$ and radius $r$ surrounded by an SCL potential dip of total radius $W$. Bohm-accelerated ions flow from the plasma into the dip with flux $\Gamma_i = n_i c_s$, where $c_s = \sqrt{T_e / m_i}$ is the sound speed. These ions occasionally collide with neutrals that have density $n_0$ and temperature $T_0 \lesssim T_s$, \latestChange{where $T_s \approx$ 1800--2200 K} is the temperature of the hot surface. The number of charge-exchange events is proportional to the cross-section of such interactions, $\sigma_{cx}$, which is known for low-temperature xenon plasmas to be around 90 $\angstrom^2$ \cite{MillerXe}. Here, it is assumed that all charge-exchange ions are sufficiently slow to be trapped in the SCL potential dip, which is usually satisfied since SCL wells are deep enough to impede electrons with energies $\approx 5 e T_s$. Altogether, the rate of accumulation for trapped ions is
\begin{equation}
\frac{dN}{dt}\bigg|_\text{accum} = n_i c_s \times \sigma_{cx} n_0 \times \pi \bigg((W+r_0)^2 - r_0^2\bigg) L.
\end{equation}

In competition with this growth rate, some loss mechanism generally drains the population of trapped ions. Here, they diffuse axially. Each end of the filament is cooler than its middle due to thermal conduction to a holder, so that emission falls and the sheath dip vanishes near the filament ends. In other plasma situations, alternative mechanisms such as ion collisions, electron heating, or plasma flow may remove trapped ions. 

To quantify the trapped ion loss rate in this experiment, we consider two annular loss regions with outer radii $W$ and inner radii $r_0$. The flux through these loss regions depends on the thermal velocity of the cold ions, $v_\text{esc}$, and the trapped ion population $n_T$, as follows:
\begin{equation}
\frac{dN}{dt}\bigg|_\text{loss} = n_T v_\text{esc} \times 2 \times \pi \bigg( (W+r_0)^2 - r_0^2\bigg).
\end{equation}

Since the loss rate in this model grows with the trapped ion density $n_T$, the trapped population will increase until the loss rate equals the accumulation rate. Thus, in equilibrium,
\begin{equation}
\frac{dN}{dt}\bigg|_\text{accum} = \frac{dN}{dt}\bigg|_\text{loss} \implies \frac{n_T}{n_i} = \frac{c_s}{v_\text{esc}}
\frac{\sigma_{cx} n_0 L}{2}.
\end{equation}
As might be expected, the equilibrium value of $n_T$ scales most strongly with the neutral density $n_0$, since uncharged species must be present in the sheath for charge-exchange events to create trapped ions.

How much will this trapped ion population increase the floating potential of the emitting surface? We seek the additional potential contribution $\phi_T$ that arises from the space charge of trapped ions, which only builds up within the potential dip at radii \mbox{$r_0 < r < r_0 + W$}. This potential is found by integrating Poisson's equation in cylindrical geometry such that \mbox{$-\nabla^2 \phi_T = en_T/\epsilon_0$} for \mbox{$r < r_0 + W$} and $\nabla^2\phi_T = 0$ otherwise (as in Fig.~\ref{fig:cartoon}c). For boundary conditions, we first choose \mbox{$\phi_T(r_0) =0$} since potentials are relative; second, we assume that the sheath-presheath electric field, \mbox{$E(r_0+S) = T_e/\lambda_D$} \cite{GodyakSternberg, Kaganovich2002}, is established entirely by ambient plasma properties and emitted electrons, and thus enforce \mbox{$E_T(r_0 +S) = -\nabla \phi_T(r_0 + S) = 0$} for the trapped ion contribution. These conditions define the potential due to trapped ions inside the dip as 
\begin{equation}
\phi_T(r) = \frac{e n_T}{2 \epsilon_0}\left[ (r_0 + W)^2 \ln\frac{r}{r_0} + \frac{r_0^2 - r^2}{2} \right],
\end{equation}
so that the total potential change at the edge of the dip $\Delta \phi_T = \phi_T(r_0 + W)$ is
\begin{equation}\label{eqn:xi}
\Delta \phi_T = \frac{e n_T r_0^2}{2\epsilon_0}\left[(1+\xi )^2 \ln \left( 1 + \xi\right)  - ( \xi + \xi^2/2 )\right],
\end{equation}
where $\xi = W /r_0$ is the width of the potential dip normalized to the object size. The expression is only true up to a certain $n_T$, after which the potential rises enough that the SCL dip disappears and ions can are no longer trapped; this saturation may or may not lead to steady-state equilibrium \cite{Campanell2017}. Our time-averaged measurements observe a marginally positive sheath, indicative of saturation around $\Delta \phi_T \gtrsim \Delta \phi_\text{SCL}$. In any case, Eq.~\ref{eqn:xi} depends strongly on $\xi$, a parameter which has not been measured for floating cylindrical or spherical objects.  

It should be noted that this model considers trapped ion space charge in isolation, ignoring any self-consistent response of electron emission. Still, high electron mobility and the overwhelming flux from a strongly emitting surface should establish the potential well equilibrium on much faster timescales than the slow ion current buildup. A full simulation that includes these feedback effects is beyond the scope of this work. Notwithstanding, $\Delta \phi_T(\xi)$ can be used to examine the feasibility of \again{cumulative ion trapping (and resulting positive sheaths)} in given plasma conditions.

The above analysis suggests that trapped ion charge did not cause the reduction of the sheath potential drop observed in Fig.~\ref{fig:noFlow}. At neutral pressures as high as \mbox{$8 \times 10^{-5}$} Torr, the trapped ion density \modify{is predicted to reach} \mbox{$n_T = 0.04\ n_i = 1.2 \times 10^{14}$ m$^{-3}$}. A uniform density this low would only compensate the usual charge of an SCL sheath, \mbox{$\Delta \phi_T(\xi) = 1.5\ T_e$}, if $\xi > 40$. Such a high value of $\xi$ is not likely when the surface scale length $r$ is already on the order of $\lambda_D$. As such, reversing this sheath potential by trapped ion buildup would require an unphysically large potential dip width.

However, plasmas with high background neutral density can support trapped ion populations that are nearly as dense as the ambient plasma. Near a high-pressure inductively coupled RF discharge \cite{GodyakIEPC}, the above accumulation model calculates a trapped ion density of $n_T = 0.85\ n_i = 2.5 \times 10^{17}$ m$^{-3}$. The filament was several centimeters from an antenna that produced ambient plasma with $n_e = n_i = 3 \times 10^{17}$ m$^{-3}$ and a near-Maxwellian electron temperature of $T_e = 4$ eV; the chamber was filled with neutral Xe gas at $p_0 \gtrsim 1$ mTorr. Eq.~\ref{eqn:xi} predicts that the higher trapped density $n_T$  would overcome the negative SCL sheath as long as $\xi \gtrsim 1$. The width of this potential dip has not been directly measured, but it is expected that \mbox{$W \gtrsim$ 2--3 $\lambda_D \approx r_0$} \cite{Intrator, Robertson}. \again{Thus, the model of ion trapping is consistent with the buildup of positively floating emitting sheaths} in this discharge.

\begin{figure}[h]
\centering
\includegraphics[width=0.5\textwidth]{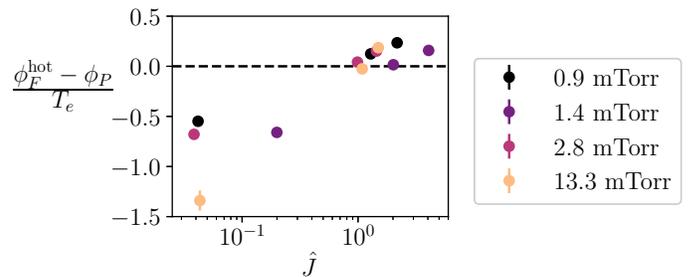}
\caption{\label{fig:RF} The emissive sheath potential drop, normalized to $T_e$ in eV, is shown in an RF plasma discharge with varied neutral pressure, plotted against the emitted current $\hat{J}$. When $\hat{J} \gtrsim 1$, the emissive filament \again{obtained a positive potential}, floating about 0.2 $T_e$ above the plasma potential. \change{Error bars include contributions from measurements of $\phi_F^\text{hot}$, $\phi_P$, and $T_e$.} \finalRevision{Absolute values of $\hat{J}$ are approximate due to exponential dependence on $T_s$ \mbox{\cite{Dushman}}, but relative values are measured with a high precision of less than 5\%.}}
\end{figure}

Measurements from the plasma described above are shown in Fig.~\ref{fig:RF}. An immersed filament was heated to different degrees of emission in four plasmas with different neutral pressures. When the normalized emission current of the hot tungsten surface reached \finalRevision{$\hat{J} \approx 1$}, a positive floating sheath with $\phi_F^\text{hot} = \phi_P + 0.2\ T_e$ was measured at all pressures. These potentials reflect a stable equilibrium plasma, since the RF-induced oscillations in $\phi_P$ have amplitude less than 1 V. These measurements only constrain the potential drop between sheath and plasma, and cannot therefore comment on the monotonicity of the sheath potential profile. Nevertheless, the measured emissive sheath potential drop is consistent with the model of ion trapping, since the surface floats several $T_s$ above $\phi_P$.

Also visible in this data is an apparent threshold in $\hat{J}$, below which the floating potential remains negative. \finalRevision{Current conservation implies that no inverse sheath forms until $\hat{J} > 1$. The inferred $T_s$ and resulting $J_\text{emit}$ are consistent with the threshold occurring at an absolute value of $\hat{J} \approx 1$, and we thus interpret the negatively floating surfaces in \mbox{Fig.~\ref{fig:RF}} as emitters too overwhelmed by $J_e$ to float positively, despite ion trapping.}

\begin{figure}[t]
\centering
\includegraphics[width=0.5\textwidth]{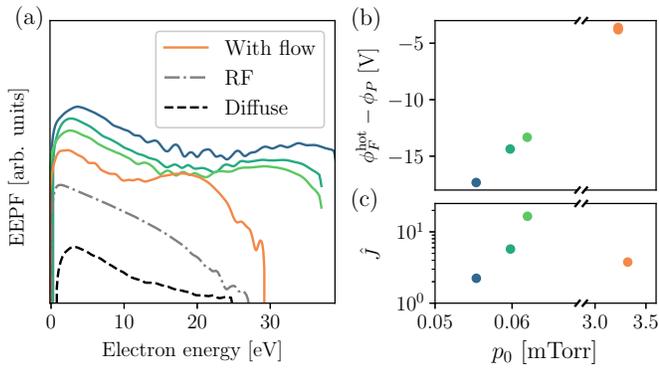}
\caption{\label{fig:EEDFs} Several EEPFs (a) from the flowing plasma discharge are shown with nonmonotonic tails. Each probability function is derived from the current-voltage characteristic of a swept probe \cite{Godyak2011}. Probability functions from the two other described plasmas without flow are shown for reference. A filament in the flowing plasma measured negative sheath potential drops (b). The current in (c) varies somewhat between measurements.}
\end{figure}

These two mechanisms \again{leading to} positive emitting sheaths, overwhelming emission and ion trapping, have been \again{considered in} unmagnetized plasmas with near-Maxwellian electron energy distribution functions. In comparison, many laboratory devices produce flowing plasmas: fast electron beams in particular are predicted to incite oscillations in the SCL sheath {\cite{Sydorenko}}, which may disrupt the buildup of trapped ions enough to prevent the surface from \again{floating at a positive potential}. Moreover, presence of flowing atoms and ions may deplete the charge-exchange ion source, while electric fields may provide sinks for trapped ions to leave the sheath. Both factors are important for measurements near Hall thrusters {\cite{Sheehan2011}}. In the final series of measurements presented here, the kHz time resolution available was too slow to measure any such dynamics, and could only register a negative time-averaged sheath potential. This plasma had $n_e = 8 \times 10^{16}$ m$^{-3}$, and the electron energy probability functions (EEPFs) shown in Fig.~\ref{fig:EEDFs}a reflect a bulk electron temperature around $T_e = 4$ eV and a variation in beam energies $E_\text{beam} > 30$ eV emitted from the cathode. As shown in Fig.~\ref{fig:EEDFs}b, the emissive sheaths in this plasma were negative, suggesting SCL sheath structures with magnitudes based not on the bulk $T_e$ but on the beam energy $E_\text{beam}$. Further studies should investigate the impacts of \latestChange{both electromagnetic fields and ion, electron and neutral} distribution functions on sheath formation, including whether flowing electrons create oscillating SCL sheaths.

As mentioned above, the object scale length $r_0/\lambda_D$ greatly affects the structure and magnitude of emissive sheaths. Even the standard SCL sheath changes quantitatively when the normalized object size $r_0/\lambda_D$ decreases, with geometrical contraction increasing near-surface ion density \cite{Fruchtman} and the width and depth of the SCL potential dip \cite{Robertson}. These same mechanisms are entangled in the \again{formation of positive sheaths through both overwhelming emission and ion trapping}: OML and contraction effects exaggerate the potential contributions of both emitted electrons and trapped ions. Systematic studies that control for emission \finalRevision{(with accurate absolute measurements of $J_\text{emit}$)} and trapping may be able to isolate these geometrical effects.

\begin{table}[t]
\normalsize
\centering
\begin{tabular}[t]{ l ||  C{2.0cm} | C{1.9cm} | C{2.2cm} } 
Parameter & Diffuse \cite{Vekselman} & RF \cite{GodyakIEPC} & With flow \cite{Fedotov} \\
\hline \hline
\rule{0pt}{3ex}    
$n_e$ [m$^{-3}$] & $3 \times 10^{15}$ & $3 \times 10^{17}$ & $8 \times 10^{16}$ \\
$T_e^\text{eff}$ [eV] & 7 & 4 & 8 \\
$\lambda_D/r_0$ & 2-3 & 0.1-0.3 & 0.6-0.8 \\
\finalRevision{$\hat{J}$ (approx.)} & 200--800 & 0.04--4 & 2--20 \\
$n_T$ [m$^{-3}$] & $10^{14}$ & $10^{17}$ & $2 \times 10^{15}$ \\
\end{tabular}
\caption{\label{table:params} Approximate parameters for each plasma environment considered.}
\end{table}

Though these observations show that the SCL sheath is not ubiquitous, it is still commonplace. Many experiments that verified the negative potential drop of emitting sheaths were performed with low emission \cite{SheehanPRL}, where monotonic sheath profiles exclude the application of either theory explained above. However, we reinforce \latestChange{that $\phi_F < \phi_P$, as suggested by standard sheath theory and the SCL model, is no universal rule}. The data indicates that electron sheaths increase the emissive sheath potential drop from $\approx T_e$ to zero, whereas \again{inverse} sheaths surround surfaces that float about $0.2\ T_e$ above $\phi_P$. The sheath structure that forms near an emitting surface depends on the ambient populations of plasma and neutral particles, and \again{a given} situation---e.g., surface charging, probe error quantification, divertor physics---may be explained by one or several models.

The experiments described here advance two modifications to the SCL model that had not been measured in the laboratory. The models, overwhelming emission and ion trapping, are shown to eliminate the negative potential drop between a floating emitting surface and a plasma. These effects must be considered to determine the insulating properties of emissive sheaths.

\begin{acknowledgements}
The authors are grateful for the experimental support of A.~Merzhevskiy, A.~Alt and I.~Romadanov, and for inspiration to perform the described experiments from A.~Khrabrov, I.~Kaganovich, and D.~Sydorenko. This work was performed under the auspices of the U.S. Department of Energy through contract DE-AC02-09CH11466.
\end{acknowledgements}

\end{document}